\documentclass[longauth,letter,traditabstract]{aa} 
\usepackage{graphicx}
\usepackage{txfonts}
\begin{document}

   \title{``TNOs are cool": A survey of the trans-neptunian region\thanks{{\em Herschel}
          is an ESA space observatory with science instruments provided
          by European-led Principal Investigator consortia and
          with important participation from NASA.}}

   \subtitle{II. The thermal lightcurve of (136108)~Haumea}

   \author{E.\ Lellouch\inst{1} \and
        C.\ Kiss\inst{2} \and
        P.\ Santos-Sanz\inst{1} \and
        T.\ G.\ M\"uller\inst{3}  \and
        S.\ Fornasier\inst{1,4} \and
        O.\ Groussin\inst{5} \and
        P.\ Lacerda\inst{6} \and
        J.\ L.\ Ortiz\inst{7} \and
        A.\ Thirouin\inst{7} \and
        A.\ Delsanti\inst{1,5} \and
        R.\ Duffard\inst{7} \and
        A.\ W.\ Harris\inst{8} \and
        F.\ Henry\inst{1} \and
        T.\ Lim\inst{9} \and
        R.\ Moreno\inst{1} \and
        M.\ Mommert\inst{8} \and
        M.\ Mueller\inst{10} \and
        S.\ Protopapa\inst{11} \and
        J.\ Stansberry\inst{12} \and
        D.\ Trilling\inst{13} \and
        E.\ Vilenius\inst{3} \and
        A.\ Barucci\inst{1} \and
        J.\ Crovisier\inst{1} \and
        A.\ Doressoundiram\inst{1} \and
        E.\ Dotto\inst{14} \and
        P.\ J.\ Guti\'errez\inst{7} \and
        O.\ Hainaut\inst{15} \and
        P.\ Hartogh\inst{11} \and
        D.\ Hestroffer\inst{16} \and
        J.\ Horner\inst{17} \and
	L.\ Jorda\inst{5} \and
        M.\ Kidger\inst{18} \and
        L.\ Lara\inst{7} \and
        M.\ Rengel\inst{11} \and
        B.\ Swinyard\inst{9} \and
        N.\ Thomas\inst{19} 
 }
 
   \institute{Observatoire de Paris, 
              5 Place Jules Janssen, 92195 Meudon Cedex, France\\ \email{emmanuel.lellouch@obspm.fr}  
\and Konkoly Observatory of the Hungarian Academy of Sciences,           
              H-1525 Budapest, P.O.Box 67, Hungary 
\and Max-Planck-Institut f\"ur extraterrestrische Physik,                
 Giessenbachstrasse, 85748 Garching, Germany;
\and Universit\'e Paris-7 ``Denis Diderot", 4 rue Elsa Morante, 75205 Paris Cedex 13 
\and  Laboratoire d'Astrophysique de Marseille, CNRS \& Universit\'e de 
Provence, 38 rue Fr\'{e}d\'{e}ric Joliot-Curie, 13388 Marseille cedex 
13, France                                                               
\and Newton Fellow of the Royal Society, Astrophysics Research Centre, Physics Building,                     
              Queen's University, Belfast, County Antrim, BT7 1NN, UK 
\and              Instituto de Astrof\'isica de Andaluc\'ia (CSIC)       
              C/ Camino Bajo de Hu\'etor, 50, 18008 Granada, Spain
\and Deutsches Zentrum f\"ur Luft- und Raumfahrt,                        
              Berlin-Adlershof, Rutherfordstra{\ss}e 2,
              12489 Berlin-Adlershof, Germany 
\and Space Science and Technology Department,                            
              Science and Technology Facilities Council,
              Rutherford Appleton Laboratory,
              Harwell Science and Innovation Campus,
              Didcot, Oxon UK, OX11 0QX
\and Observatoire de la C\^ote d'Azur, laboratoire Cassiop\'ee           
              B.P. 4229; 06304 NICE Cedex 4; France 
\and          Max-Planck-Institut f\"ur Sonnensystemforschung,           
              Max-Planck-Stra{\ss}e 2, 37191 Katlenburg-Lindau, Germany 
\and Stewart Observatory,                                                
              The University of Arizona, Tucson AZ 85721, USA 
\and         Northern Arizona University, Department of Physics \& Astronomy,    
              PO Box 6010, Flagstaff, AZ 86011, USA 
\and INAF--Osservatorio Astronomico di Roma, Via di Frascati, 33,              
              00040 Monte Porzio Catone, Italy  
\and              ESO, Karl-Schwarzschild-Str.\ 2,                                    
              85748 Garching bei M\"uchen, Germany 
\and IMCCE, Observatoire de Paris, 77 av. Denfert-Rochereau, 75014 Paris 
\and          Department of Physics and Astronomy, Science Laboratories, Durham University, South 
Road, Durham, DH1 3LE, UK   
\and              European Space Agency (ESA),                                        
              European Space Astronomy Centre (ESAC),
              Camino bajo del Castillo, s/n,
              Urbanizacion Villafranca del Castillo,
              Villanueva de la Ca\~nada,
              28692 Madrid, Spain 
\and          Universit\"at Bern, Hochschulstrasse 4,                             
              CH-3012 Bern, Switzerland }

   \date{Revised \today; accepted}

 
  \abstract{Thermal emission from Kuiper Belt object (136108)~Haumea was measured with {\em Herschel}--PACS at 100~$\mu$m and 160~$\mu$m  
for almost a full rotation period. Observations clearly indicate a 100~$\mu$m thermal lightcurve with an amplitude of
a factor of $\sim$ 2, which is positively correlated with the optical lightcurve. This confirms that both are primarily
due to shape effects. A 160~$\mu$m lightcurve is marginally detected. Radiometric fits of the mean {\em Herschel-} and {\em Spitzer-}
fluxes indicate an equivalent diameter D$\sim$1300 km and a geometric albedo p$_v$$\sim$0.70-0.75. These values agree  with
inferences from the optical lightcurve, supporting the hydrostatic equilibrium hypothesis.  The large amplitude of the 100~$\mu$m lightcurve
suggests that the object has a high projected a/b axis ratio  ($\sim$1.3) and a low thermal inertia as well as possible variable infrared beaming. This may point to fine regolith on the surface, with a lunar-type photometric behavior. The quality of the thermal data is not sufficient to clearly detect the effects of a surface dark spot.     }


   \maketitle
%

\section{Introduction}
The dwarf planet (136108) Haumea (formerly 2003 EL$_{61}$) is one of the most 
remarkable transneptunian objects (TNOs). Its large amplitude 
visible lightcurve indicates a very short ($\sim$3.91 h) rotation period and considerable rotational deformation, with semi-major axes estimated to be 1000$\times$800$\times$500 km (Rabinowitz et al. 2006). It is one of the bluest TNOs (Tegler et al. 2007), and unlike other 1000 km-scale TNOs, its surface is covered by almost pure water ice (Trujillo et al. 2007), 
 though its high density 
($\sim$2.6 g cm$^{-3}$, Lacerda and Jewitt 2007) indicates a more rocky interior. It possesses two satellites (Brown et al. 2006), the larger of which 
is also water-ice coated (Barkume et al. 2006). All this, and the observation that several TNOs with similar orbital parameters
also show evidence of surface water ice (Schaller and Brown 2008), point to Haumea being the largest remnant of a massive ancient ($>$1 Gyr) collision
(Ragozzine and Brown 2007). High time-resolution, multi-color, photometry provides evidence for a surface feature redder and darker than the surrounding materials (Lacerda et al. 2008, Lacerda 2009), perhaps of collisional origin, and makes Haumea the second TNO (after Pluto) with surface
heterogeneity. Both {\em Spitzer} thermal observations (Stansberry et al. 2008) and visible photometry 
indicate that Haumea is one of the most reflective TNOs %
(estimated geometric albedo 0.6 -- 0.85).

Except for near-Earth asteroids (e.g. Harris et al. 1998, M\"uller et al. 2005), thermal lightcurves of airless bodies are available only for a few objects. 
In combination with the optical lightcurve, they provide a means to distinguish between the effects of shape and surface 
markings and to infer thermal properties 
of the surface. Lockwood and Brown (2009) reported on the detection
of Haumea's lightcurve at 70 $\mu$m with {\em Spitzer}. We present here observations of Haumea at 100 and 160 $\mu$m with {\em Herschel}-PACS, 
performed in the framework of the Open Time Key Program ``TNOs are cool" (M\"uller et al. 2009, 2010).

%


\section{{\em Herschel} / PACS observations }
\subsection{Observations and data reduction}
Haumea was observed with the PACS photometer (Poglitsch et al.\ \cite{poglitsch10}) of {\em Herschel}  (Pilbratt
et al. 2010) on 2009, December 23 and 25 (Obs. ID \# 1342188470
and \# 1342188520, respectively), using the 100 $\mu$m (``green") / 160 $\mu$m (``red") combination. We used
a mini scan-map mode  which homogeneously covered a field roughly 1\,arcmin in diameter
(M\"uller et al. 2010). 

Although our observations on December 23 were initially designed to cover 110 \% of Haumea's visible lightcurve, 
they lasted only 3.36 hr (i.e. 86 \% of the 3.91 hr period) end-to-end
(UT 5:52:01--9:13:25) due to shorter than expected observations overheads in the mini-scan mode. Observations on December 25 
lasted only 40 minutes (UT 6:13:39--6:53:59). Their goal was 
to verify the measured
target flux at a given rotational phase against a different sky background (Haumea moved by about 85''
between these two dates).

The whole observation sequence for December 23 consisted of 400 individual measurement
subscans. 
As a compromise between temporal resolution and sensitivity, we divided
the sequence into ten, $\sim$20-min long blocks (40 subscans each) for the 100 $\mu$m data and into five, $\sim$40-min 
blocks (80 subscans each) at 160 $\mu$m. Similarly, the 40 individual subscans of December 25 were combined
into two blocks at 100 $\mu$m and one single block at 160 $\mu$m. 
Individual measurements were reduced 
with standard scan map processing, 
applying masked high pass filtering in the
vicinity of bright sources (only Haumea was notably bright in the central
region of the maps) and resampling maps with pixel sizes of 1'' in the green and 2'' in the red. 
For the high-pass filter we used widths of 15'' and 20" for
the green and red bands, respectively. The selection criterion for the mask 
was set to be the 3~$\sigma$ value of the image.
The calibration was done in a standard way, applying flux 
overestimation corrections of 1.09 and 1.29 at 100 and 160 $\mu$m, respectively, as 
recommended 
at the time of the processing (PICC-ME-TN-036, 22-Feb-2010,
see {\it herschel.esac.esa.int}). 
Color corrections (to monochromatic reference wavelengths of 100.0 and 160.0 $\mu$m) are at the 1 \% level, i.e. negligible
in view of other uncertainties. 


\vspace*{-.5cm}
\begin{figure}[ht]
\centering
\includegraphics[width=8.5cm,angle=0]{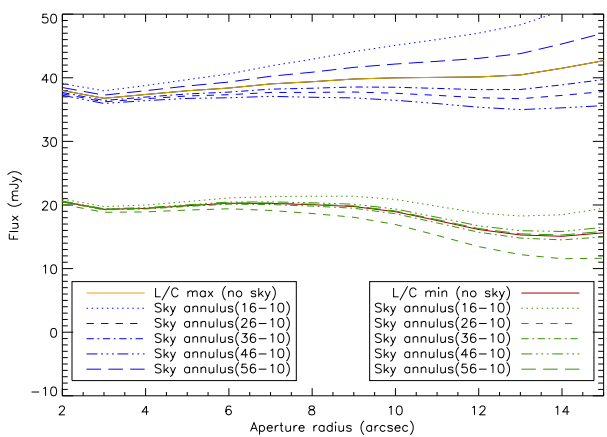}
\caption{Photometry and associated uncertainties in 100-$\mu$m band. Aperture-corrected curves-of-growth
are shown for the two visits of Dec. 23 corresponding to lightcurve maximum and minimum. The orange
and red curves are for measurements uncorrected for sky contribution. The other curves are for
sky-corrected measurements, with the sky contribution estimated in annuli centered on Haumea
and with various internal and external radii (r$_{int}$ = 16-56'' and  r$_{ext}$ = r$_{int}$ + 10'').}
\label{fig:photom}
\end{figure}

\vspace*{-.8cm}
\subsection{Photometry}
For each visit to Haumea, we performed standard aperture photometry with the 
IRAF/Daophot flux extraction routines and aperture correction technique (Howell 1989). 
We constructed photometric curves of growth, using aperture radii ranging from 1'' to 15'',
and performed aperture corrections based on tables of
the fraction of the encircled energy of a point source. 
The optimum aperture for photometry was selected from inspection of the curves of growth,
and was usually found to be 1.0-1.25 times the PSF FWHM (7.7'' in the green and 12'' in the red) in radius,
and to lie in the ``plateau" zone of the curves of growth.
In determining the ``optimized" source flux in this manner, we did not subtract any sky contribution,
as the latter is normally eliminated in the data reduction process. However,
to assess the uncertainty on this flux value, we also constructed sky-subtracted curve-of-growths, 
selecting a variety of regions of the image (typically annuli centered on the source) to measure the 
sky contribution. This method is illustrated in Fig. \ref{fig:photom} for two of the 100 $\mu$m visits to Haumea,
corresponding to the maximum and minimum fluxes. 
The standard deviation in all flux values determined in this manner and for a broad 
range of aperture radii (3''--15'') finally provided the 1$\sigma$ uncertainties 
attached to the flux measurements. 

\section{Phasing with visible observations}
We observed 
Haumea in the visible on 2010 January 20, 21, 23, and 26, using  a 0.4~m f/3.5 telescope located in San 
Pedro de Atacama (Chile) and equipped with a 4008 x 2672 CCD camera. A broad band filter 
(390-700nm) was used with integration times of 300~s. Additional data were
acquired on 2010 January 28, with the 1.2~m telescope at Calar Alto Observatory (Spain),
equipped with a 2k$\times$2k CCD camera in the R filter, again with 300~s integration times. 
The same reference stars were consistently observed each night. 
Observations were reduced and analyzed as in Ortiz et al. (2007), with 
some refinements described in Thirouin et al. (2010). The 3.92 hr period, 0.28 mag
amplitude lightcurve was readily detected, and by combining these data with
the Lacerda et al. (2008) 
observations,  an improved rotation period of 3.915341$\pm$0.000005 h was derived. This allowed us to
phase the January 2010 observations back to the time of the {\em Herschel} observations
very accurately.

\section{Results and analysis}

\subsection{Thermal lightcurve}
\begin{figure}[ht]
\centering
\includegraphics[width=9cm,angle=0]{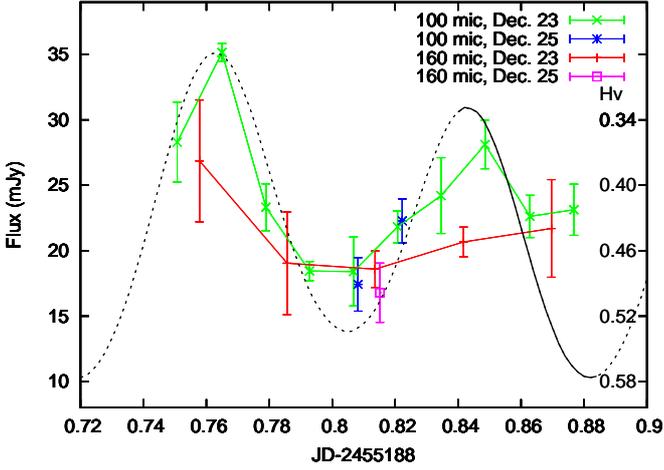}
\caption{Observed thermal lightcurve of Haumea at 100 $\mu$m (green) and 160 $\mu$m (red).
Black dots show the visible lightcurve (H$_V$, right scale); the enhanced part corresponds to the
fraction affected by the presence of the dark spot in the Lacerda et al. (2008) model.}
\label{fig:lc}
\end{figure}

The measured Haumea fluxes on Dec. 23 (10 points at 100 $\mu$m, 5 points at 160 $\mu$m) 
are plotted in Fig. \ref{fig:lc} 
as a function of fractional date. At 100 $\mu$m a clear lightcurve is detected, with mean $\sim$25 mJy flux
and high contrast (18-35 mJy, i.e. almost a factor of 2 peak-to-peak), while a more marginal lightcurve is determined 
at 160 $\mu$m. Superimposed on Fig. \ref{fig:lc} are the fluxes measured on Dec. 25 (2 points at 100 $\mu$m, 1 point at 160 $\mu$m),
rephased to Dec. 23 using the 3.915341 h period. Their agreement with the Dec. 23 measurements demonstrates
the robustness of the lightcurve. To further verify the large amplitude and overall structure of the 100 $\mu$m
lightcurve, we also performed  ``differential photometry", subtracting the median or averages of all combined images of Dec. 23 from each individual image. 
This confirmed a peak-to-peak lightcurve amplitude of $\sim$17 mJy in the green.

Overall, the 100 $\mu$m lightcurve appears positively correlated with the visible lightcurve, as expected if 
shape effects are dominant. Nonetheless, the secondary peak (near JD2455188.843) and absolute
minimum (JD2455188.883) of the visible lightcurve, attributed to the
presence of a dark spot (Lacerda et al. 2008) are also associated with
a secondary maximum and minimum in the thermal lightcurve. If
anything, the dark spot should be warmer than the rest of the surface,
and therefore tend to enhance the thermal emission. This is observed
for the secondary minimum but the secondary peak shows the opposite
behavior. We qualitatively conclude that the thermal lightcurve
confirms the elongated shape of Haumea, but does not unambiguously
support the presence of a spot.

\subsection{Radiometric size and albedo}
We first performed radiometric modeling of the mean fluxes, combining the mean 100 and 160 $\mu$m
fluxes (25$\pm$2 mJy and 21$\pm$3 mJy, respectively) with {\em Spitzer} results at 24 and 70 $\mu$m (Stansberry et al. 2008).
The latter indicate a color-corrected flux of 13.4$\pm$2.0 mJy at 71.42 $\mu$m and upper limit of 0.025 mJy at 23.68 $\mu$m,
for measurements performed on 2005 June 22, UT = 9:11-9:40, roughly mid-way between visible lightcurve maximum and minimum.
(Note that the 7.7 mJy value published in Stansberry et al. (2008) is the average of the above
with a much lower flux (2.5 mJy) measured on 2005 June 20; we suspect this
latter observation was compromised).

\begin{figure}[ht]
\centering
\includegraphics[width=9cm,angle=0]{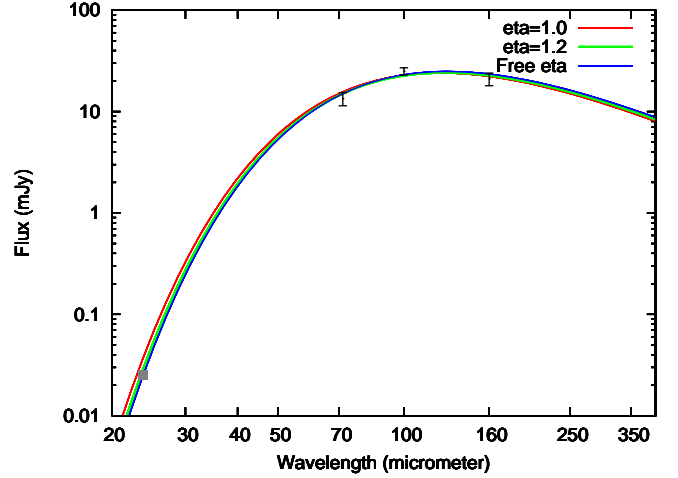}
\caption{Approximate Haumea SED measured from {\em Herschel}/PACS (100 and 160 $\mu$m) and {\em Spitzer} (24 and 70 $\mu$m).
The grey square at 23.68 $\mu$m is the {\em Spitzer} upper limit. Model fits using NEATM with fixed $\eta$ ($\eta$ = 1.0
and $\eta$ = 1.2 are shown) and free $\eta$ (best fit $\eta$ = 1.38) are shown.}
\label{fig:radiom}
\end{figure}

Modeling was performed along the NEATM approach (Harris et al., 1998, M\"uller et al. 2010), a.k.a. the ``hybrid-STM" (Stansberry et al. 2008). 
Essentially, the temperature distribution across the object follows instantaneous equilibrium with local insolation,
but is modified by an empirical factor $\eta^{-1/4}$, where $\eta$, the beaming parameter, can be either specified or fit to the data. In this framework
$\eta$ values much higher than 1 indicate large thermal inertia effects, while $\eta$$\leq$1 points to low thermal inertia
and important surface roughness. Free parameters
are the mean radiometric diameter D, geometric albedo p$_v$, and possibly $\eta$. We used a mean H$_R$ = 0.09 and V--R = 0.335,
i.e a mean H$_V$ = 0.425. We adopted a  phase integral q~=~0.7, intermediate between those estimated for Pluto
(0.8) and Charon (0.6) (Lellouch et al. 2000), and 
an emissivity $\epsilon$~=~0.9. We considered three cases: 
$\eta$ = 1, $\eta$ = 1.2 (mean value inferred for TNOs by Stansberry et al. 2008), and free $\eta$. Table \ref{tab:fits} gives the radiometric solution for the three cases, and 
Fig. \ref{fig:radiom} shows the associated fits. A satisfactory match of the 70, 100 and 160 $\mu$m fluxes is achieved in all
cases, though it is noteworthy that (i) when $\eta$ is a fitting parameter, it is poorly constrained  and (ii) the predicted 24 $\mu$m is always 
essentially at the {\em Spitzer} upper limit.  Note that using the ``default" q value for TNOs (0.39) would lead to p$_v$ $\sim$1, which is strongly
at odds with the observed correlation between p$_v$ and q (Lellouch et al. 2000).

\begin{table}
     \caption{Radiometric fits for (136108) Haumea }
     \label{tab:fits}
     \begin{tabular}{lcccc}
\hline
\hline
\noalign{\smallskip}
 Model & $\eta$ & D(km) & p$_v$ & Reduced $\chi^2$ \\
\noalign{\smallskip}
\hline
\noalign{\smallskip}
Fixed $\eta$ & 1.0 & 1230$\pm$18 & 0.810$\pm$0.024 & 1.14 \\ 
Fixed $\eta$ & 1.2 & 1276$\pm$20 &  0.752$\pm$0.024 & 1.08 \\
Free $\eta$ & 1.38$\pm$0.71   & 1324$\pm$167 & 0.698 $\pm$0.189 & 1.06\\
\noalign{\smallskip}
\hline
     \end{tabular}
\end{table}


\subsection{Modeling of the thermal lightcurve}

%

\begin{figure*} [ht]
\includegraphics[width=7.2cm]{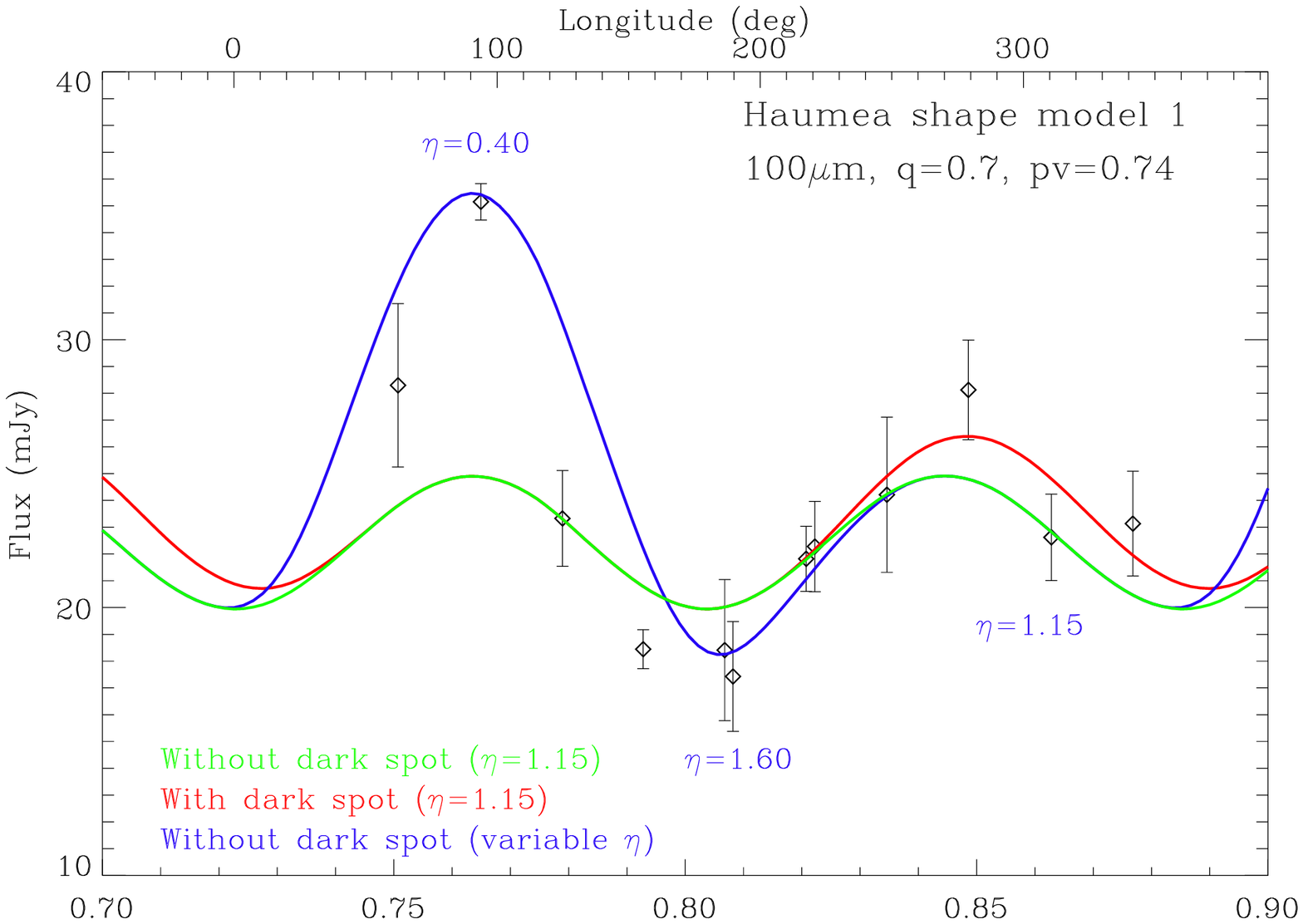}
\includegraphics[width=7.2cm]{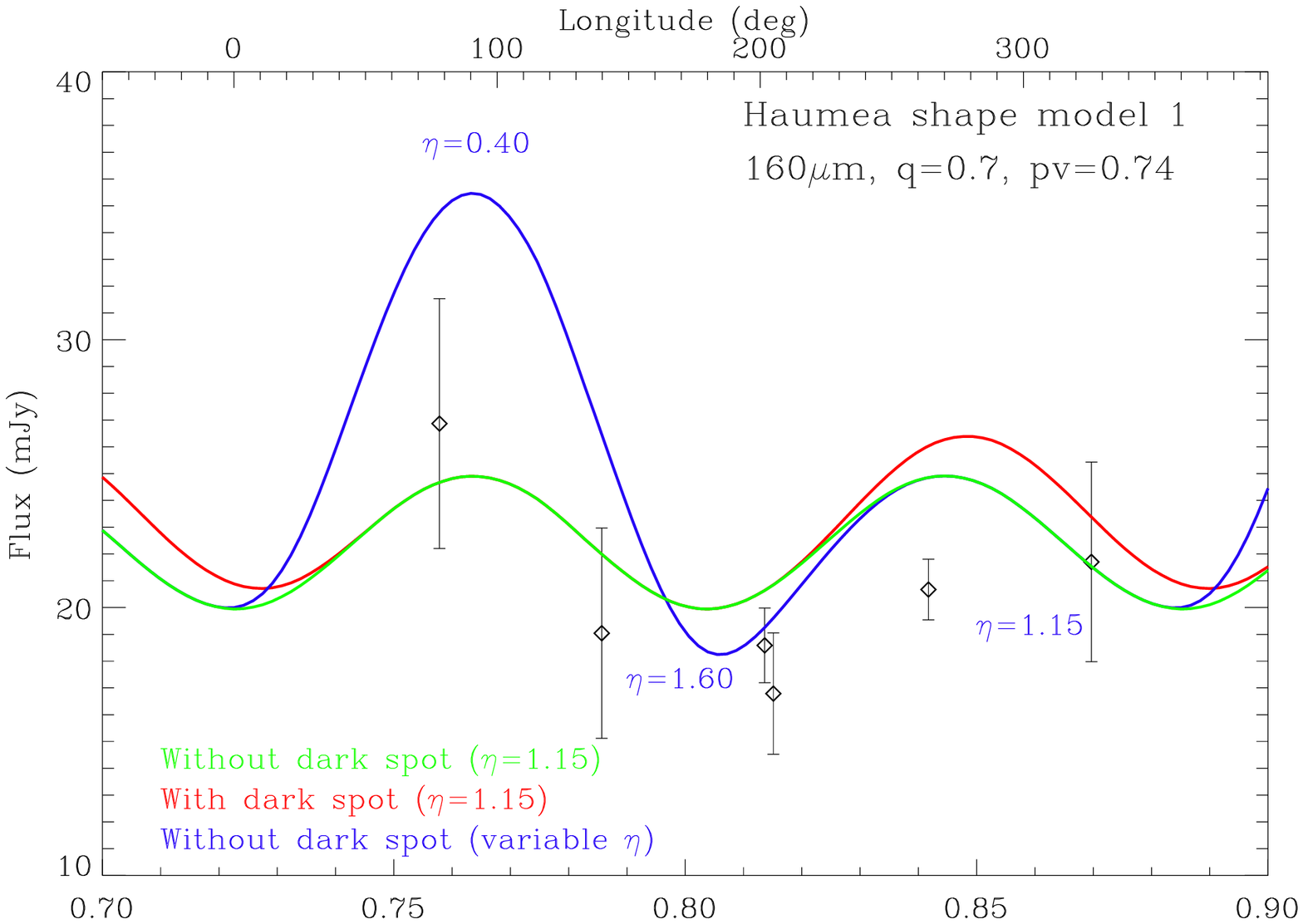}\\
\includegraphics[width=7.2cm]{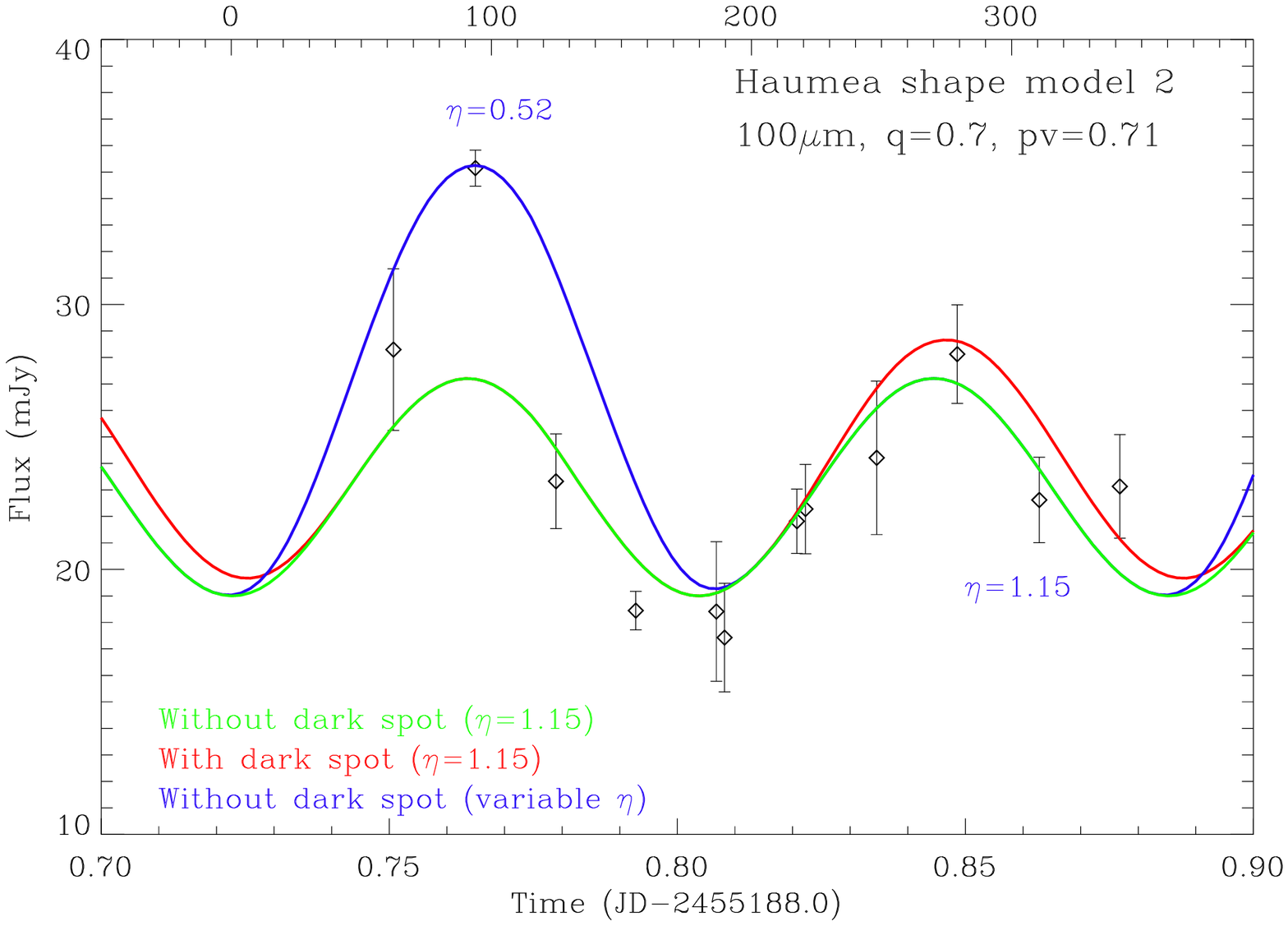}
\includegraphics[width=7.2cm]{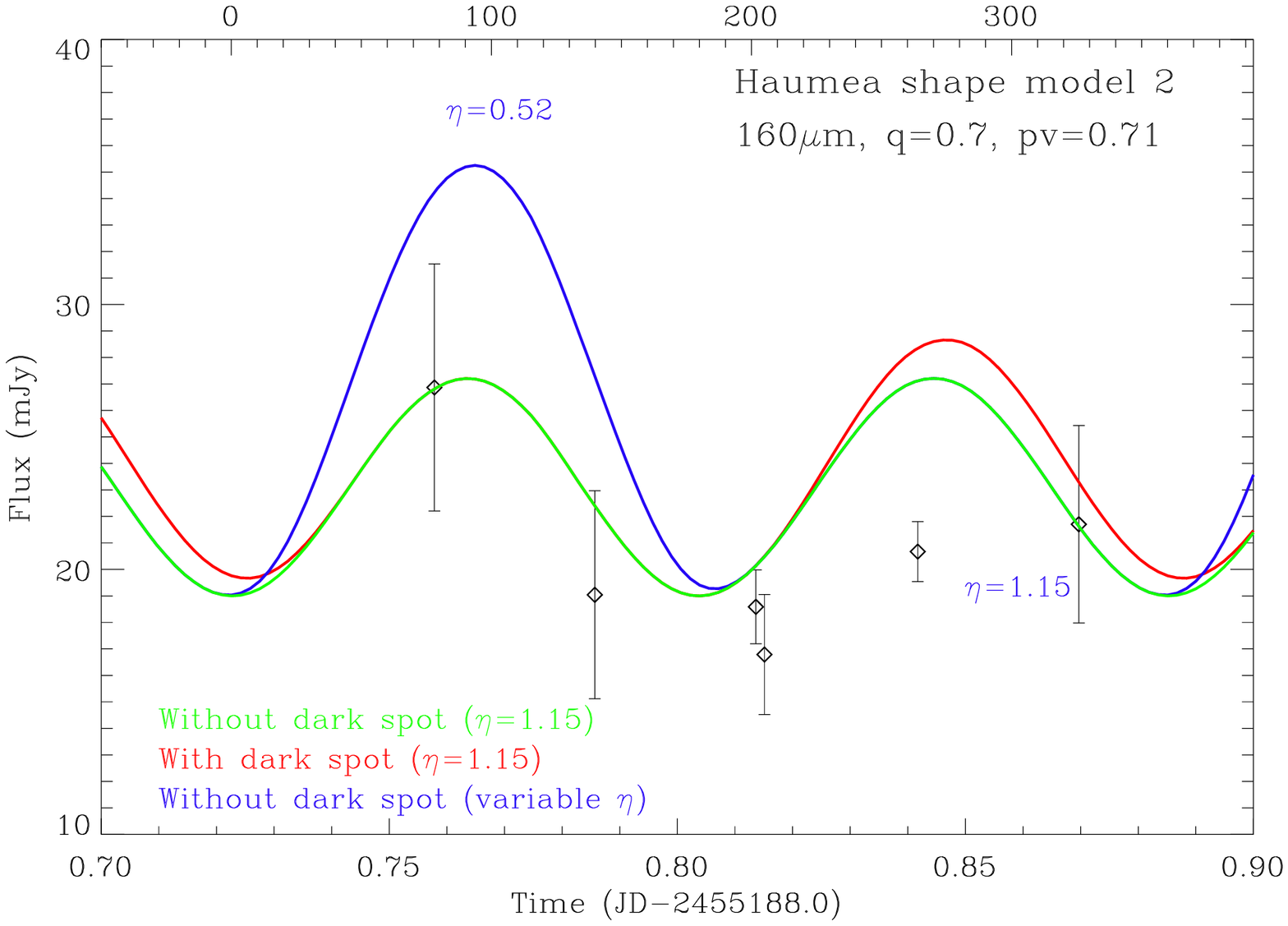}
\caption{Fits of Haumea 100 $\mu$m and 160 $\mu$m lightcurves with shape models 1 and 2 (see text) and different distributions 
of $\eta$. Green curves: constant $\eta$ (1.15, for both models). Blue curves: Variable $\eta$ as a function
of longitude $L$. For model 1, $\eta$ = 0.40 at $L$ = 80--130, $\eta$ = 1.6 at $L$ = 130--180, and $\eta$ = 1.15 elsewhere.
For model 2, $\eta$ = 0.52 at $L$ = 130--180 and $\eta$~=~1.15 elsewhere. Red curves: constant $\eta$, but a dark spot is
included. The spot covers longitudes from 292.5$^{\circ}$ to 337.5$^{\circ}$, i.e. 1/4 of one hemisphere, and its albedo is taken as 0.59 (model 1)
or 0.56 (model 2). The adopted reference longitude system is shown, in which 0$^{\circ}$ is the semi-minor axis closest to the spot.}
\label{}
\end{figure*}

To model the 100~$\mu$m and 160 $\mu$m lightcurves, the above model was modified to account
for Haumea's elongated shape. For that purpose we used a versatile
tool called OASIS, the Optimized Astrophysical Simulator for 
Imaging Systems (Jorda et al. 2010), describing the object as an
ellipsoid made of 5120 triangles. OASIS then calculates
the orientation of each triangle relative to (i)  pole orientation, (ii)  observer 
position, (iii) Sun position, and (iv) time as the object rotates. The large amplitude of Haumea's lightcurve favors a large aspect angle. We nominally used OASIS assuming
an equator-on object (aspect angle $\theta$~=~90$^{\circ}$) and with the observer at the Sun (phase angle = 0$^{\circ}$),
but given the orbits of Haumea's satellites (Ragozzine and Brown 2009), we also considered $\theta$~=~75$^{\circ}$.
The direction of the 
Sun dictated the local insolation, which was then fed into the NEATM model. Input parameters for Haumea 
were the three semi-major axes (a, b, c) and the geometric albedo (p$_v$). Those were derived following Rabinowitz et al. (2006) 
and Lacerda and Jewitt (2007), but using the measurements of Lacerda et al. (2008). Essentially the amplitude and
period of the visible lightcurve, and the assumption of a Jacobi figure (hydrostatic equilibrium) provide b/a, c/a and the density $\rho$.
Knowledge of Haumea's mass (4.0$\times$10$^{21}$ kg; Ragozzine and Brown 2009) then provides the absolute semi-major axes. Finally, p$_v$ is deduced from
H$_V$. 
 
We first used the preferred photometric solution of Lacerda et al. (2008), in which a Lambert scattering law is assumed, expected for high-albedo 
icy surfaces. For $\theta$~=~90$^{\circ}$, this yields b/a = 0.87, c/a = 0.54, and $\rho$ = 2.55 g cm$^{-3}$. In this case,  a = 927 km, b = 807 km, c = 501 km,
and p$_v$ = 0.74 (model 1). 
We also used an alternative model with a higher axial ratio, as inferred for a lunar-type Lommel-Seelinger reflectance function.
In this case, and still with $\theta$~=~90$^{\circ}$, b/a = 0.80, c/a = 0.52, $\rho$= 2.59, a = 961 km, b = 768 km, 
c = 499 km, and p$_v$ = 0.71 (model 2). We emphasize that the equivalent mean diameter, which can be taken as 
2a$^{1/4}$b$^{1/4}$c$^{1/2}$ is 1309--1317 km, in excellent agreement with the above radiometric fits; the same comment applies
for p$_v$ (see Table 1). This tends to support the hydrostatic equilibrium hypothesis. Once the object dimensions and albedo are fixed, the only free parameter
for the thermal model is $\eta$. 

For $\theta$~=90$^{\circ}$, Fig. 4 shows that model 1 with  $\eta$ = 1.15 reproduces the mean flux level, but the amplitude of the 100 $\mu$m lightcurve is  grossly 
underestimated. Model 2, again with $\eta$~=~1.15, provides a good fit to the 100~$\mu$m data, especially in the region of the secondary 
maximum near JD2455188.843, but the $\sim$35 mJy main peak near JD2455188.763 is still underpredicted. 
We explored models with spatially variable $\eta$ to try and fit this peak. These models required an extended region with 
much lower $\eta$ ($\eta$ = 0.40 and 0.52 for models 1 and 2, respectively).
For model 1, fitting the overall lightcurve structure would even require a three-terrain $\eta$ model. $\eta$ values below 0.6 imply
r.m.s. slopes well in excess of 40$^{\circ}$ (Spencer 1990, Lagerros, 1998) and seem unrealistic for this object. Yet our result might qualitatively
point to a highly craterized region with extremely strong beaming, given the very high surface albedo. Further thermophysical modeling,
including a more realistic description of surface roughness, is required. Note finally that models with $\theta$~=~75$^{\circ}$ gave similar
results as $\theta$~=~90$^{\circ}$, provided slightly higher $\eta$ values were used (e.g. $\eta$ = 1.35 instead of 1.15 for the uniform model).

%

We finally investigated the effect of a dark spot on the surface at the location inferred
by Lacerda et al. (2008) (i.e. centered at 315$^{\circ}$ in the longitude system of Fig. 4). It was prescribed to cover 1/4 of Haumea's maximum projected cross section, 
with a relative albedo constrained by  
the Lacerda et al. (their Fig. 7) results. Other combinations of spot coverage / relative albedo did not qualitatively change 
the results. As expected, the dark spot enhances the thermal flux in the relevant longitude ranges, but our data are insufficient
to demonstrate its effect.

Besides the determination of Haumea's diameter and albedo, the essential finding is that the large amplitude of the thermal lightcurve implies
both a high a/b ratio ($\sim$1.3) and a low $\eta$ ($<$1.15-1.35, depending on spin orientation) value, indicative of a low thermal inertia. This seems inconsistent
with compact water ice at $\sim$40 K, and rather points to a porous surface. As a high a/b ratio also implies a lunar-like photometric
behavior, this could point to Haumea's surface being covered with loose regolith with poor thermal conductivity. The same conclusion
is reached for other objects in our sample, such as the Plutino
2003 AZ$_{84}$ and Centaur 42355 Typhon (M\"uller et al. 2010).  Surface regolith may be 
produced by collision events and retained on the surface of large TNOs.


%
%




\end{document}